\documentclass[journal]{IEEEtran}
\usepackage{cite}
\usepackage{graphicx}
\graphicspath{{./fig/}}
\usepackage{subfigure}
\usepackage{multirow}
\usepackage{flushend}
\usepackage{algorithm}
\usepackage{algorithmic}
\usepackage{float}

\begin{document}

\title{Image Credibility Analysis with Effective Domain Transferred Deep Networks}

\author{Zhiwei~Jin,~Juan~Cao,~Jiebo~Luo,~\IEEEmembership{Fellow,~IEEE} and~Yongdong~Zhang,~\IEEEmembership{Senior Member,~IEEE}%

\thanks{This work was supported in part by the National Nature Science Foundation of China (61571424) and Beijing Advanced Innovation Center for Imaging Technology (BAICIT-2016009).}%
\thanks{Zhiwei Jin, Juan Cao, and Yongdong Zhang are with the Key Laboratory of Intelligent Information Processing of Chinese Academy of Sciences (CAS), Institute of Computing Technology, CAS, Beijing 100190, China. E-mail: \{jinzhiwei, caojuan, zhyd\}@ict.ac.cn.}%
\thanks{Jiebo Luo is with the Department of Computer Science, University of Rochester, Rochester, NY 14627, USA. E-mail: jluo@cs.rochester.edu.}}

\maketitle

\begin{abstract}
Numerous fake images spread on social media today and can severely jeopardize the credibility of online content to public. In this paper, we employ deep networks to learn distinct fake image related features. In contrast to authentic images, fake images tend to be eye-catching and visually striking. Compared with traditional visual recognition tasks, it is extremely challenging to understand these psychologically triggered visual patterns in fake images. Traditional general image classification datasets, such as ImageNet set, are designed for feature learning at the object level but are not suitable for learning the hyper-features that would be required by image credibility analysis. In order to overcome the scarcity of training samples of fake images, we first construct a large-scale auxiliary dataset indirectly related to this task. This auxiliary dataset contains 0.6 million weakly-labeled fake and real images collected automatically from social media. Through an AdaBoost-like transfer learning algorithm, we train a CNN model with a few instances in the target training set and 0.6 million images in the collected auxiliary set. This learning algorithm is able to leverage knowledge from the auxiliary set and gradually transfer it to the target task. Experiments on a real-world testing set show that our proposed domain transferred CNN model outperforms several competing baselines. It obtains superiror results over transfer learning methods based on the general ImageNet set. Moreover, case studies show that our proposed method reveals some interesting patterns for distinguishing fake and authentic images.

\end{abstract}


\section{Introduction}
\IEEEPARstart{R}{ecent} years have seen the rapid growth of many online social media platforms, such as Facebook, Twitter and its Chinese equivalent Weibo\footnote{http://www.weibo.com/}. Users not only acquire daily information generated on social media, but also propagate and even contribute new content on the networks. The large amount of user generated data is valuable for opinion mining and decision making. However, the convenience of posting content on social media also fosters numerous fake information. Without analyzing the credibility, fake content would spread rapidly through social networks and result in serious consequences \cite{friggeri2014rumor} \cite{zhao2015enquiring}, especially in the case of emergency events, such as ``\textit{Hurricane Sandy}" \cite{gupta2013faking} and ``\textit{Malaysia Flight MH370 Lost Contact}" \cite{jin2014news}.

\begin{figure}[!t]
\centering
\subfigure[Outdated images]{

\includegraphics[width=2.8in]{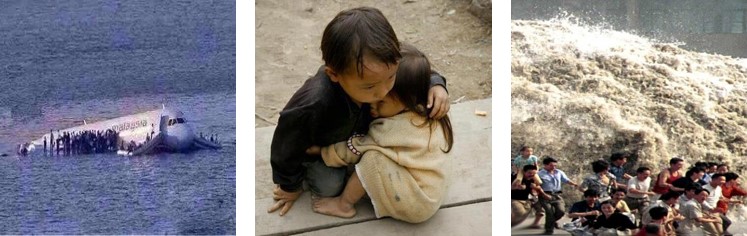}
\label{fig_1a}
}
\subfigure[Inaccurate images]{
\includegraphics[width=2.8in]{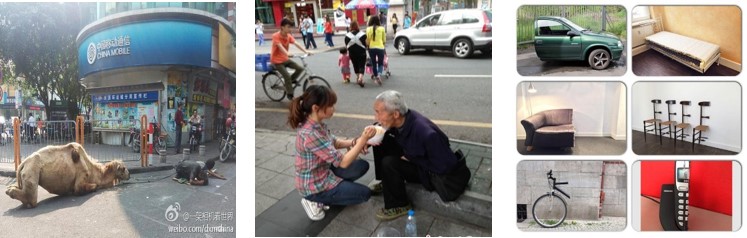}
\label{fig_1b}
}
\subfigure[Manipulated images]{

\includegraphics[width=2.8in]{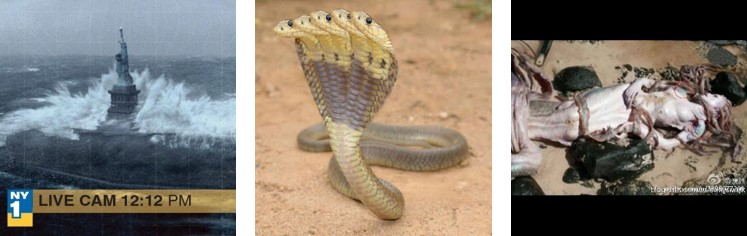}
\label{fig_1c}
}
\caption{Different types of fake images: (a) outdated images used for current events; (b) images used intentionally to support false events; (c) digitally manipulated images.}
\label{fig_1}
\end{figure}

In this paper, we aim to visually analyze the credibility of images. Most existing studies focus on the textual content to analyze the credibility of online content. Based on features extracted from surrounding text or social context, classification-based methods \cite{castillo2011information} \cite{kwon2013prominent} \cite{wu2015false} \cite{jin2015} and graph-based optimization methods \cite{gupta2012evaluating} \cite{jin2014news} \cite{jin2016} \cite{zhou2015} are deployed to identify posts as fake or authentic. Only a few recent studies focus on verifying the credibility of multimedia content \cite{gupta2013faking} \cite{jin2015} and they are still based on the surrounding textual features of images.

Compared with pure texts, images are vivid and easily comprehensible. They are prevalent on social media and can rapidly reach a huge number of people. Many recent studies, including sentiment analysis \cite{you2015robust} and opinion mining \cite{fang2015}, have integrated visual content for better understanding of online content.

Recently, the visual credibility analysis problem has received much attention in the research and industry area. Boididou \emph{et al.} [3] propose the Verifying Multimedia Use task which took place as part of the MediaEval benchmark in 2015 and 2016. This task attracts
many researchers to analyze the credibility of images and videos on Twitter. Pantti and Sir{\'e}n \cite{Mervi} explore possible approaches to use non-professional images on social media for journalism. They start a project\footnote{http://www.verified-pixel.com/} to verify user generated images by combining human examination and algorithm detection.

\begin{figure}[!t]
\centering
\includegraphics[width=3.3in]{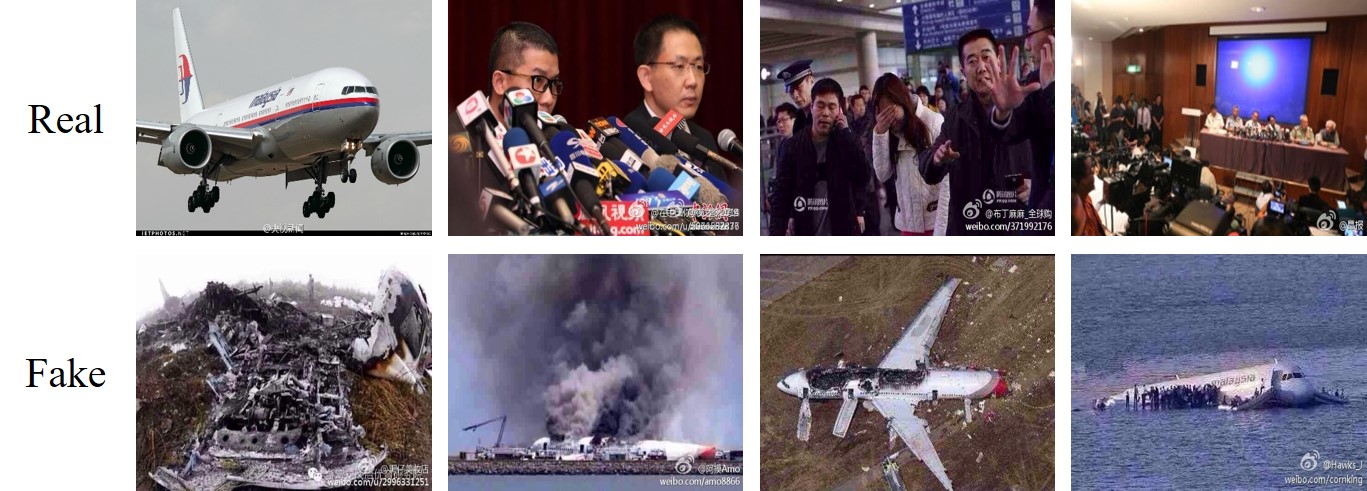}
\caption{Some images in ``\textit{Malaysia Flight MH370 Lost Contact}". Real images are in the top row and fake images are in the bottom row. Despite the difference in appearance, fake images tend to be eye-catching and visually striking compared with real images.}
\label{fig_2}
\end{figure}

Boididou \emph{et al.} \cite{Bodidou2015Challenges} define fake images as images in posts that do not accurately represent the events that they refer to. To be more specific, we define fake images as images attached in fake tweets.\footnote{The truthfulness of tweets are determined by authoritative sources in this paper. Thus, images would have authoritative labels with this definition.} Many different types of fake images exist on social media: they may be outdated images associated with current events, digitally manipulated images, artworks or other fabricated images referring to a current event. For example, we show three different types of fake images in Figure \ref{fig_1}.

\begin{itemize}
  \item
  \textbf{Outdated images} Figure \ref{fig_1a} shows three outdated images used in current events. The left image showing MH370 plane crushed into sea is actually an accident happened in New York in 2009. The middle one claimed to be two children in 2015 Nepal earthquake is a picture of two Vietnamese taken in 2007. The right image used to report Indian Ocean tsunami is actually the scene of the tidal bore of Chinese Qiantang River.
  \item
  \textbf{Inaccurate images}  Figure \ref{fig_1b} shows cases where images are inaccurately used to describe false events. The left image claims a camel with limbs cut off used for begging is fake because the camel is actually resting with legs bent under itself. The middle image claimed as a kind girl helping a homeless old man turns out to be a posed picture for commercial exploration. The right image claimed as an angry woman breaking her husband's things into halves is also a commercial event.
  \item
  \textbf{Manipulated images} Figure \ref{fig_1c} illustrates digitally tampered images. The huge storm, six-head snake and mermaid are all elaborately edited images.
\end{itemize}

Different types of images could be fake under certain circumstances. It is very challenging to deal with such variations in visual appearance. However, we can still find some common patterns of fake images from the psychological angle. From the fake and real images referring to the accident ``\textit{Malaysia Flight MH370 Lost Contact}" in Figure \ref{fig_2}, we can observe that fake images tend to be eye-catching and visually striking in contrast to real ones. Certainly, these psychologically triggered patterns are beyond direct visual appearance and much more complicated than common object-level features in traditional visual recognition tasks. Therefor, a traditional image set, such as the ILSVRC-2012 dataset of the ImageNet challenge, is not suitable for our task.

\begin{figure}
\centering
\includegraphics[width=3.5in]{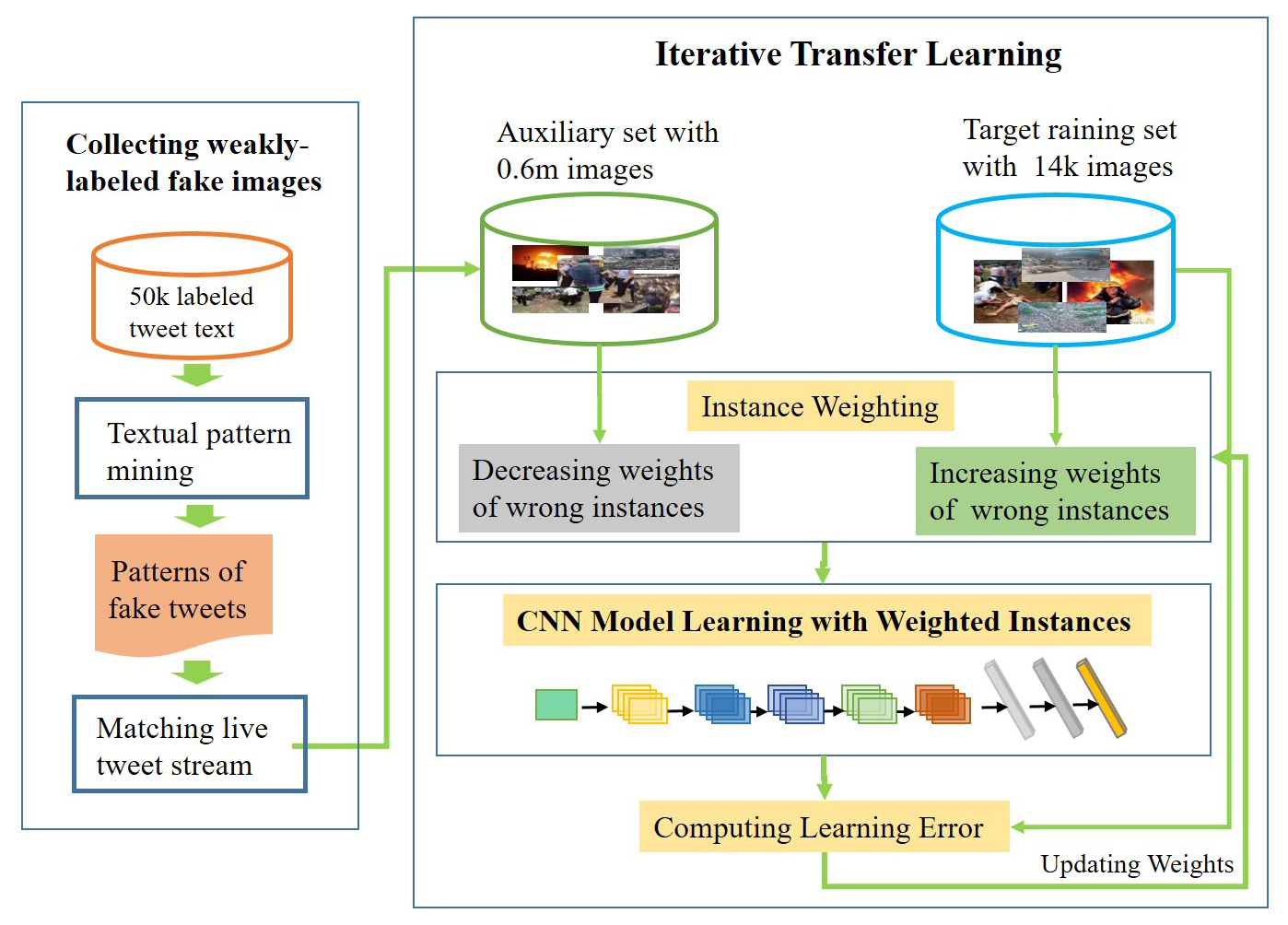}
\caption{Our proposed approach for image credibility analysis consists of three main procedures: (1) Constructing the auxiliary image set for transfer learning. (2) Training a CNN model with weighted instances from both the auxiliary set and target training set. (3) Transfer learning: knowledge is transferred from the auxiliary domain to the target task by weighting instances iteratively.}

\label{fig_3}
\end{figure}

Directly constructing a dataset for image credibility analysis requires massive labor of image annotation by human experts. In addition, training a CNN to learn complicated fake image features requires a large amount of labeled images. In this paper, we propose a domain transferred CNN to overcome data deficit and capture complicated features for image credibility analysis. In particular, our approach has three main procedures:

First, a large-scale image set related to our task is constructed for transfer learning. To overcome insufficient training samples for training a reliable CNN model, we resort to transfer learning. Existing general image sets are not suitable for image credibility analysis task. We propose to construct an auxiliary set directly related to our task. After mining the textual patterns of fake tweets, we harvest millions of weakly-labeled fake images from social media automatically to form a related auxiliary set for transfer learning.

Second, fake image related features are learned with CNN. Convolutional neural networks (CNN) have demonstrated outstanding ability in learning robust and accurate image representations. We employ a deep CNN to learn complicated features for detecting fake images. General CNN models assume instances have equal contribution to the final result. However, in the process of transfer learning with the auxiliary set and target set, some images in the auxiliary set may have misleading impact on the final target task. We assign different weights to images accordingly and build a CNN to learn with weighted instances.

Finally, we transfer knowledge learned in the auxiliary set to image credibility analysis task through an iterative transfer learning algorithm. Due to the distributional difference between the auxiliary set and target training set, some instances in auxiliary set may be noisy and can degrade the final performance. We adopt a transfer learning algorithm with instance weighting similar to \cite{Dai2007boost}. It iteratively decrease the impact of ``bad instances'' and gradually transfer knowledge to the target domain. With some labeled samples from the target domain, our domain transferred CNN can produce accurate and robust results. The framework of our proposed image credibility analysis approach is illustrated in Figure \ref{fig_3}.

Our main contributions are summarized as follows.

\begin{itemize}
  \item
Unlike existing studies based on textual content, we directly exploit visual content for image credibility analysis by implementing a deep convolutional neural network to learn representations of images. This CNN model can effectively capture complicated characteristics of fake images and outperform traditional handcrafted textual or visual features.
  \item
We construct a large-scale image set for the image credibility analysis task. Based on textual patterns detected from fake tweets, we have collected a set of  over 600,000 weakly-labeled fake and real images from social media.
  \item
We propose to train a CNN model with both collected auxiliary set and target training set through transfer learning. An AdaBoost-like instance weighting algorithm is adopted to distinguish the more useful instances from the less useful ones in the auxiliary set. The CNN model is also specifically designed to deal with weighted instances during learning.
  \item
With the knowledge transferred from auxiliary set and the learning ability of CNN, we obtain a robust image recognition model for image credibility analysis. Experimental results show that our approach produces promising performance and outperforms several competing baseline methods.

\end{itemize}

\section{Related Work}
In this section, we provide a brief review of research most closely related to our work in two aspects: credibility analysis of online information, and CNN-based transfer learning.

\subsection{Credibility Analysis of Online Information}
Fake information spreading on social networks has become a serious concern as social media platforms gain huge popularity around the world \cite{morris2012tweeting}. To detect fake online information, most existing studies focus on the text content. Supervised classification methods are widely used to identify fake online posts. In \cite{castillo2011information}, \cite{kwon2013prominent}, and \cite{wu2015false}, the authors detect false rumor posts on Twitter or Weibo with features extracted from the text content, user, topic, and propagation process. Rather than classifying each message individually, the propagation-based approaches \cite{gupta2012evaluating} are presented to analyze information credibility as a whole. Inter-event relations \cite{jin2014news} or conflicting relations among messages \cite{jin2016} are exploited for propagating credibility among messages. Some works propose to improve the performance of text credibility analysis with features related to attached images. Morris \emph{et al.} \cite{morris2012tweeting} release a survey declaring that user profile image has great impact on the credibility of posts published by this user. Similarly, Gupta \emph{et al.} \cite{gupta2012evaluating} define a feature to record whether a user has a profile image or not, and Wu \emph{et al.} \cite{wu2015false} define a ``has multimedia"  feature to indicate the status of multimedia attachment of a tweet.

Only a few recent studies directly addresse image credibility analysis. Gupta \emph{et al.} \cite{gupta2013faking} make an effort to understand the temporal, social reputation and influence patterns for the spreading of fake images on Twitter. They propose a classification model to identify fake images during ``\emph{Hurricane Sandy}". Their work is still based on traditional text and user features. Aiming to automatically predict whether a tweet that shares multimedia content is fake or trustworthy, Boididou \emph{et al.} \cite{verifying2015} propose the Verifying Multimedia Use task as part of the 2015 MediaEval benchmark. This task attracts attentions for analyzing the credibility of images in tweets. None of these studies explore the deep learning features due to the lack of sufficient labeled training data.

\subsection{CNN-based Transfer Learning}
Analyzing image credibility requires an accurate comprehension of image semantics. Recently, Convolutional Neural Network (CNN) has shown a promising result on learning image representation and capturing semantic meaning of images. For many computer vision problems, including image classification \cite{krizhevsky2012imagenet} \cite{simonyan2014very} and object detection \cite{girshick2014rich} \cite{szegedy2013deep}, CNN has shown its superiority to traditional models based on hand-crafted features.

We intend to use CNN for learning complicated semantics in fake images. A typical CNN contains several convolutional layers and several fully connected layers. A deep CNN requires millions of parameters learned during the supervised training process. For example, The AlexNet in \cite{krizhevsky2012imagenet} contains more than 60 million parameters. It is almost impossible to train this network with a few thousand images. To overcome this limitation, we resolve to transfer learning in this paper.

Transfer learning aims to transfer knowledge from the auxiliary domain to a related target domain \cite{pan2010survey}. In computer vision, some works \cite{tommasi2010safety} \cite{aytar2011tabula} use transfer learning to overcome the deficit of training samples. They adapt classifiers trained with sufficient samples for other categories to the target category. Some recent works propose transferring image representations learned from the large-scale ImageNet data set using the CNN architecture of \cite{krizhevsky2012imagenet}. They investigate transferring CNN features to traditional visual recognition tasks such as Caltech256 image classification \cite{zeiler2014visualizing}, scene classification \cite{donahue2013decaf}, object recognition \cite{girshick2014rich} \cite{sermanet2013overfeat} \cite{oquab2014learning}. More similar to our work, \cite{you2015robust} trains a deep network with a large-scale weakly labeled dataset for visual sentiment analysis. To enable transfer learning, it requires a task-related auxiliary set and a proper transfer learning algorithm. In this paper, we propose to collect task-related images through a text mining procedure and iteratively transfer from the auxiliary domain to the target domain through an instance weighting algorithm.

\section{Image Credibility Analysis}

\begin{figure*}
\centering
\includegraphics[width=6.5in]{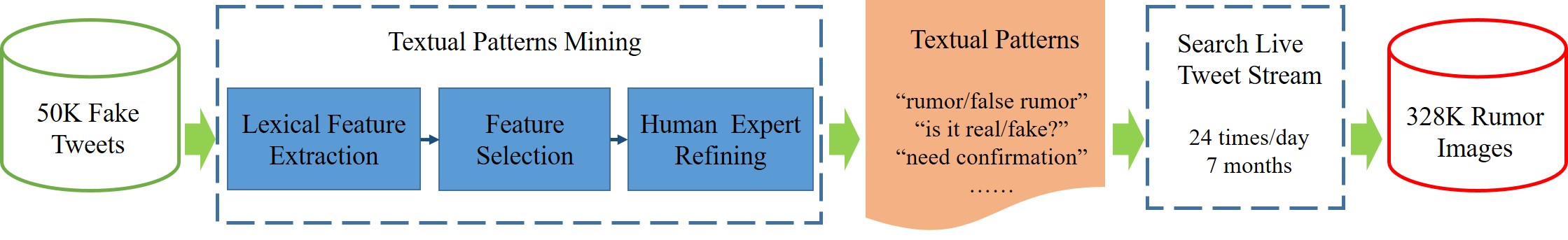}
\caption{The pipeline for automatic construction of an auxiliary image set.}
\label{fig_4}
\end{figure*}

Our proposed framework of image credibility analysis with domain transferred CNN has three main procedures. Firstly, we construct an auxiliary image set with a large amount of weakly-labeled fake images. An automatic method for harvesting these image from social media is proposed based on text mining of fake tweets. Secondly, a Convolutional Neural Network is trained with the auxiliary set as well as the  target training set. The network is designed to train with weighted instances. Finally, an AdaBoost-like instance weighting algorithm is adopted for dealing with the distributional difference issue of auxiliary set. We give details of these procedures in following sections.

\subsection{Automatic Construction of Auxiliary Image Set}

\subsubsection{Challenges in Auxiliary Set Construction}

Most existing CNN-based transfer learning are trained with general image sets, such as ILSVRC-2012 dataset for image classification tasks in ImageNet Challenge. Image credibility analysis can also be treated as a classification problem, but it has more complicated problem settings than common image classification tasks. To determine the credibility of images requires understanding of complicated semantics of image content. Models trained on traditional image sets would perform poorly for the image credibility analysis. For reliable transfer learning, it requires an auxiliary image set directly related to our image credibility analysis task.

In order to train an accurate CNN with transfer learning, it is extremely crucial to construct a large-scale auxiliary image set directly related to the image credibility analysis task. Each image in this set should be labeled to indicate its credibility polarity. Moreover, this set should contain a large amount of such images so that it can cover the huge variation among various events and supply sufficient samples for learning millions of parameters in the deep CNN. Manually annotating millions of images as fake or real demands a team of domain experts and poses extremely huge cost. It is nontrivial to obtain such an auxiliary image set.

\subsubsection{Collecting a Large-scale Auxiliary Image Set}

Surrounding textual content of images are regarded as weak guidance during the learning process in some visual recognition tasks recently. For example, tags on Flickr images are utilized as indicators of image sentiment for analyzing their sentiment polarity \cite{borth2013sentibank}. With the popularity of online social media, it is easy to collect numerous raw images along with the corresponding texts.

Inspired by the idea of using surrounding text as guidance, we propose to discover rumor patterns in textual posts via text mining. With detected patterns, we crawl a large set of matched posts from the live stream of Weibo. Attached images in these posts are then gathered to form the auxiliary image set. The pipeline of this method is illustrated in Figure \ref{fig_4}.

\begin{figure}[!t]
\centering
\subfigure[Weakly-labeled real images]{

\includegraphics[width=3in]{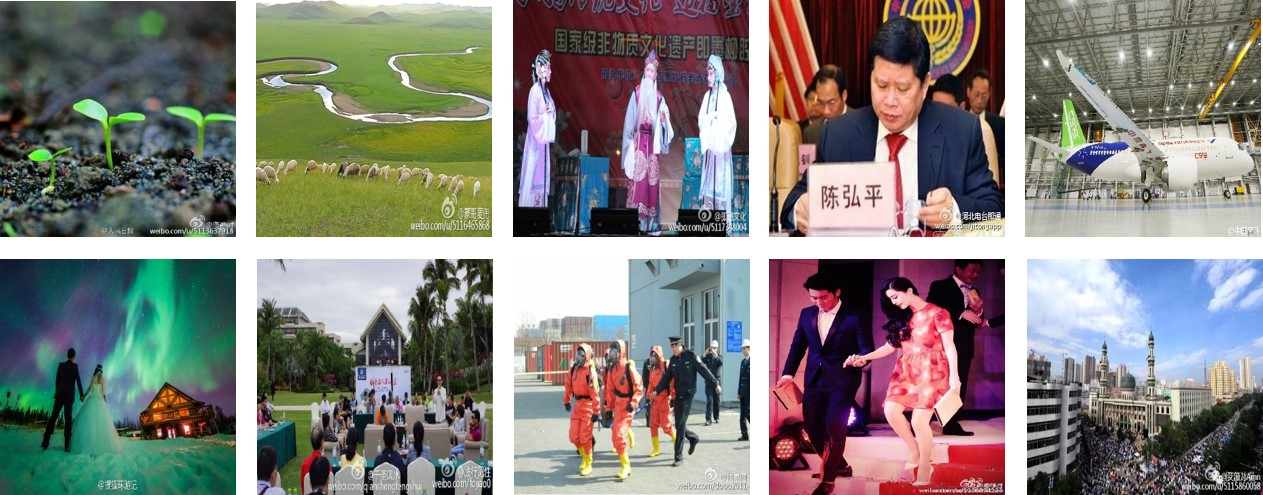}
\label{fig_5a}
}
\subfigure[Weakly-labeled fake images]{
\includegraphics[width=3in]{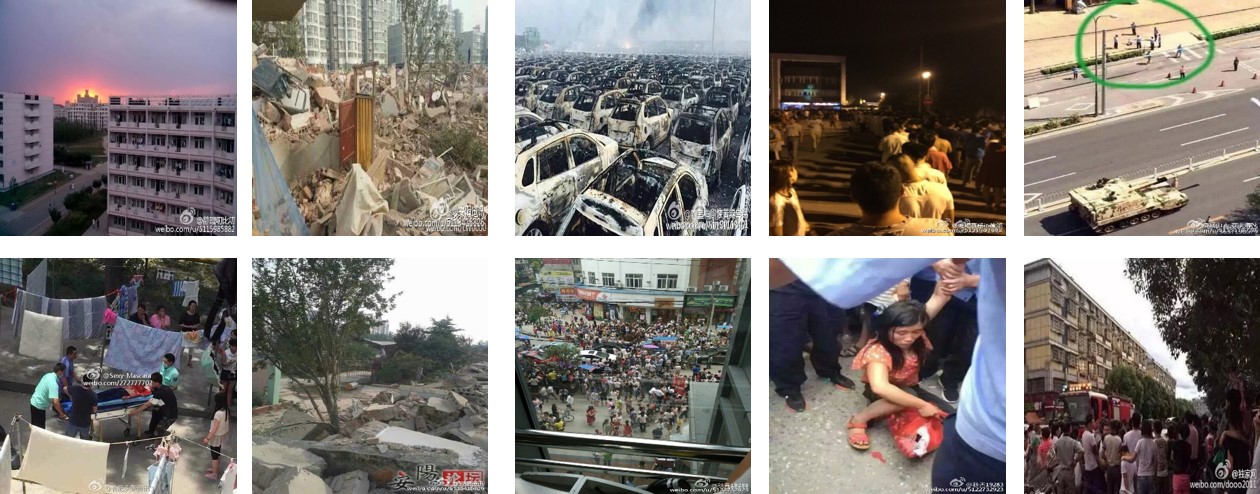}
\label{fig_5b}
}

\caption{Some fake and real images in the auxiliary set. It is clear that fake images in this set also very eye-catching.}
\label{fig_5}
\end{figure}

\textbf{Mining Textual Patterns of Fake Tweets}
The key element of our method is mining reliable patterns representing fake tweets. According to the study of \cite{zhao2015enquiring},
the informative part of fake tweets can be of the following two types: inquiries of the event situation (verification/confirmation questions) and corrections/disputes of the event. We aim to detect the inquiry and correction patterns in fake tweets through supervised feature selection on a set of labeled fake tweets.

We conduct analysis on the fake news dataset presented in \cite{jin2016}. This set contains 23,456 fake tweets and a similar number of real tweets from 146 different fake and real news events collected from Weibo. Uni-grams, bi-grams, and tri-grams lexical parts are extracted from these tweets after standard word segmentation. Frequency features (tf) of each extracted words and phrases are then calculated. We then apply the Chi-Squared test \cite{forman2003extensive} and information gain ratio method \cite{yang1997comparative} to select prominent patterns. The feature selection methods rank features based on their ability to distinguish fake tweets from real ones. From the ranked list of features, human experts select event-independent phrases as final patterns for fake tweets. Finally, we obtain these patterns, including verification patterns ( ``is it real/fake?", ``need more evidence/confirmation", \emph{etc.}) and correction patterns ( ``rumor/false rumor", ``spreading/ fabricating rumor", \emph{etc.}).

These patterns are employed for crawling live tweet streams. Specifically, we construct queries with detected patterns as keywords and search the tweet stream with the Weibo Search Engine\footnote{http://s.weibo.com/}. We run this crawling algorithm for about seven months from July 2015 to January 2016. The search interval is one hour. After crawling about one million fake tweets, we harvest images attached to them. We then remove duplicated images and finally obtain a total of 328,380 weakly-labeled fake images. According to the experimental results in \cite{zhao2015enquiring}, tweets filtered with similar ``rumor patterns" are much more likely to be fake tweets. This weakly-labeled fake image set is directly related to our task and is adequate to serve as the auxiliary set for transfer learning.

To fill the auxiliary set with negative samples, we crawl about an equal number of images from authoritative sources, such as official news agencies, as real images. Finally, we obtain a set of 0.6 million weakly-labeled fake and real images (Figure \ref{fig_5}), which would be sufficient for the CNN training.

\subsection{Training CNN with Weighted Instances}

For learning with images from auxiliary set and target training set, we use a similar network architecture to that of AlexNet \cite{krizhevsky2012imagenet}, which is the most typical CNN architecture in image classification area. Our network is composed of five successive convolutional layers ($C1-C5$) followed by three fully connected layers ($FC6-FC8$). The first, second and fifth convolutional layers are followed by max-pooling layers and normalization layers. The setting of convolution kernels, stride and padding are also similar to that of AlexNet. The first two fully connected layers ($FC6, FC7$) compute with ``RELU" non-linear activation functions:

\begin{equation}
{{\bf{Y}}_k} = f ({{\bf{W}}_k}{{\bf{Y}}_{k - 1}} + {{\bf{B}}_k}),\  k = 6,7.
\end{equation}

\begin{equation}
f ({\bf{X}})[i] = max(0,{\bf{X}}[i])
\end{equation}

And the last fully connected layer computes with ``SoftMax" function:

\begin{equation}
{{\bf{Y}}_8} = \sigma ({{\bf{W}}_8}{{\bf{Y}}_{7}} + {{\bf{B}}_8})
\end{equation}

\begin{equation}
\sigma ({\bf{X}})[i] = \frac{{{e^{{\bf{X}}[i]}}}}{{\sum\nolimits_j {{e^{{\bf{X}}[j]}}} }}
\end{equation}

Here, ${\bf{Y}}_k$ is the output of the $k$-th layer, ${\bf{W}}_k$ and ${\bf{B}}_k$ are the parameters of the $k$-th layer, $f ({\bf{X}})$ is the ``ReLU" function and $\sigma ({\bf{X}})$ is the ``SoftMax" function.

Unlike AlexNet where the output layer has 1000 dimensions, our network connects to a 2-dimensional output layer. Most existing CNN-based transfer learning assume that each training instance has equal contribution to the final model. However, instances from the auxiliary set only have machine-generated weak labels and some ``bad instances'' would decrease the result of target task. So we assign weights to images to indicate their importance on the target task. In order to learn with these weighted image instances, this network is trained by maximizing the following modified conditional log likelihood function:

\begin{equation}
L(\theta) = \sum\limits_{i = 1}^n {w_i[{y_i}\ln p({y_i} = 1|{x_i},\theta) + } (1 - {y_i})\ln p({y_i} = 0|{x_i},\theta)]
\end{equation}

here $x_i$ is the feature vector and $y_i=\{0,1\}$ is the label of the $i$-th instance, $n$ is the number of all instances, $\theta$ are parameters to be learned and $w_i \geq 0$ is the weight assigned to the $i$-th  instance. In the next section, we propose an iterative algorithm to compute weights for instances based on the learning error of the learned model.

\subsection{Iterative Transfer learning}
The knowledge in our collected 0.6 million images in the auxiliary set is undoubtedly valuable. With this auxiliary set, our image credibility analysis task is to learn a CNN model by leveraging the weakly-labeled instances in auxiliary and the gold-labeled instances in the target training set. With the representation learning ability of CNN, images in both sets can be embedded in the same feature space. A straightforward approach to utilize the auxiliary set is to simply combine it with the target training set and treat them equally during the CNN training. However, since distributional differences always exist between the auxiliary and target sets and some labels of auxiliary instances may be inaccurate, a classifier learned from this simple combination would not necessarily achieve a better performance. Sometimes the noise in the auxiliary set may cause the model to predict incorrectly on the test instances from the target set.

To address the distributional difference and label inaccuracy issue, we treat image instances in two sets differently. We assign each instance a weight to indicate its importance and adjust the weight accordingly during model learning. For those image instances in the auxiliary set $T_a$ that are more similar to the image instances in the target training set $T_t$, we should assign them larger weights to emphasize their importance; conversely, for those instances in $T_a$ that are less similar to the instances in $T_t$, we should assign smaller weights to them to weaken their impacts.

As shown in Algorithm 1, we adopt an iterative algorithm to updates instance weights according to the performance of a basic CNN model $P_t$ in each iteration. This algorithm is very similar to traditional AdaBoost method where the accuracy of a basic learner is gradually boosted by adjusting instance weights. In particular, we updates weights from two aspects: the weights of wrongly-predicted instances in auxiliary set $T_a$ is decreased and the weights of wrongly-predicted instances in target training set $T_t$ is increased. For an image $x$, $P_t(x) = \{0,1\}$ is the predicted label sign for $x$ and $L(x)$ is its label sign in the set. For any auxiliary image instance $x_a \in T_a$, its weight is decreased by a factor of $\beta^{|P_t(x_a)-L(x_a)|} \in (0,1]$ if it is misclassified. Thus, in the next iteration these misclassified auxiliary instances, which are dissimilar to the target instances, will affect the learning process less than the current round. Conversely, for any target image instance $x_t \in T_t$, its weight is increased by a factor of $\beta_t^{-|P_t(x_t)-L(x_t)|} \in [1,+\infty)$ if it is misclassified. Thus, these misclassified target instances will have larger impact in the next iteration. Following these weight updating rules, the weights of auxiliary instances would never larger than that of target instances, which indicates that the target set would always have larger influence on in CNN training. According to the analysis in \cite{freund1997} and \cite{Dai2007boost}, this algorithm minimizes both the error on the target set and the weighted average loss on the auxiliary set simultaneously and the training loss will gradually converges to zero with the iteration times increasing.

\begin{figure}
  \rule{\linewidth}{1pt}
  Algorithm 1: Iterative Transfer Learning \\
  \rule{\linewidth}{.5pt}
  \begin{algorithmic}
    \STATE \textbf{Input}: auxiliary set $T_a$, target training set $T_t$, and iteration number $K$.
    \STATE \textbf{Initialize}: Let $n=|T_a|$, $m=|T_t|$. Set the initial weight vector ${\bf{w}}^1=\{w_1^1,...,w_n^1,w_{n+1}^1,...,w_{n+m}^1\}$.
    \FOR {$t=1,...K$}
    \STATE 1. Set ${\bf{p}}^t={\bf{w}}^t/(\sum_{i=1}^{n+m}{w_i^t})$.
    \STATE 2. Train an instance-weighted CNN model $P_t$ with the combined training set $T_a \bigcup T_t$ and the weight distribution ${\bf{w}}^t$.
    \STATE 3. Calculate the learning error of $P_t$ on $T_t$: \[\epsilon_t = \frac{\sum_{i=n+1}^{n+m}{w_i^t \cdot |P_t(x_i)-L(x_i)| }}{\sum_{i=n+1}^{n+m}{w_i^t}}\]
    \STATE 4. Set $\beta_t=\frac{\epsilon_t}{1-\epsilon_t}$, $\beta=\frac{1}{1+\sqrt {2\ln {n/K}}}$.
    \STATE 5. Update weight vector ${\bf{w}}^t$:
    \[ w_i^{t+1} = \left\{ \begin{array}{l} w_i^t\beta^{|P_t(x_i)-L(x_i)|},  1\leq i \leq n  \\
    w_i^t\beta_t^{-|P_t(x_i)-L(x_i)|},  n+1 \leq i \leq n+m \end{array} \right. \]
    \ENDFOR
    \STATE \textbf{Output}:
    \[ P(x) = \left\{ \begin{array}{l} 1,  if \sum\limits_{i = 1}^n{\log{\frac{1}{\beta_t}}P_t(x)} \geq \sum\limits_{i = 1}^n{\frac{1}{2}\log{\frac{1}{\beta_t}}}  \\
    0,  otherwise \end{array} \right. \]
  \end{algorithmic}
  \rule{\linewidth}{1pt}
\end{figure}

\section{Experiments}
In this section, we first present a dataset with gold standard labels as the training and testing set for the image credibility analysis task. We conduct extensive experiments on this set to validate the effectiveness of our proposed approach. After describing experimental setup and baseline methods, we compare the performance of different methods for the image credibility analysis task.

\subsection{Target Image Set}
In order to provide a fair evaluation on our proposed method, we collect a dataset for the image credibility analysis task from Weibo. Compared with the auxiliary image set, images in this dataset have objective ground-truth labels. Specifically, we crawl all the fake posts (from May, 2012 to January, 2016) on the official rumor reporting system of Sina Weibo\footnote{http://service.account.weibo.com/}. This system encourages common users to report suspicious tweets on Weibo. A committee composed of reputable users then will judge the cases and verify them as fake or real. It is a very authoritative source to collect fake tweets \cite{wu2015false}. The hot news detection system of Xinhua News Agency, which is the most authoritative news agency in China, is used as the real tweet source. The raw set contains about 40k images. Images on social media are usually redundant and noisy. We remove redundant images from the raw set with a duplicated image detection algorithm based on locality sensitive hashing (LSH) \cite{Slaney2008}. We also remove very small or long images to maintain a high quality set. After balancing the real and fake images, we obtain a total of 14,616 images. The details of both the target and auxiliary sets are listed in Table \ref{table_1}. The original tweet text of the images are also crawled during this process for further experiments. We split this target image set into the training and testing sets with a ratio of 9:1 for the following experimental evaluation.

\begin{table}
\centering
\caption{Details of the Target and Auxiliary Image Sets}
\label{table_1}
\begin{tabular}{|l|c|c|c|c|}
\hline
 & Fake & Real & All & Label\\
\hline
Target Set & 7,308 & 7,308 & 14,616 & authoritatively labled\\
\hline
Auxiliary Set & 328,380 & 300,856 & 629,236 & weakly labled\\
\hline
\end{tabular}
\end{table}

\subsection{Experimental Setup}
For training the CNN model in our method, we adopt the implementation based on Caffe \cite{jia2014caffe}. During the training process, each image is resized to 224$\times$224. Images are also horizontally and vertically flipped. The dropout rate is 0.5 for the first two fully connected layers in each network. Both networks are trained on a single GeForce Tesla K20 with 5GB memory.

For training only with the auxiliary set, the learning rate starts from 0.01 for all layers. It changes to 0.001 and then 0.0001 until the error rate stops declining. It takes about 3 days to fully train the network. For fine-tune training the network, parameters in the five convolution layers and the first two fully connected layers are initialized with the corresponding parameters in the original network. The learning rate of the last layer ($FC8'$) is set to 10 times larger than that of other layers.

\subsubsection{Parameter Transfer}
As illustrated in Figure \ref{fig_3}, we design a very similar network for the image credibility analysis task. While labels in the source task is weakly labeled and noisy, labels in the target task all have confident ground-truth. The training data for the target task is very limited. The fully-supervised deep network will generally overfit the training data. In order to train a robust classifier, we transfer weights learned in the source task to the target task. We remove the output layer $FC8$ of the pre-trained network and add an adaptation layer $FC8'$ as replacement. The parameters of five convolutional layers $C1-C5$, and two fully connected layers $FC6$ and $FC7$ are first trained on the source task, then transferred to the target task. We then train the new network with confidently labeled data in the target task.

\begin{table}
\centering
\caption{Features for Text-based Method}
\label{table_2}
\begin{tabular}{|l|}
\hline
 \multicolumn{1}{|c|}{ \textbf{Feature}} \\
\hline
Number of exclamation/question mark  \\
\hline
Number of positive/negative words\\
\hline
Sentiment score \\
\hline
Number of words/characters \\
\hline
Number of first/second/third order of person \\
\hline
Number of People/Location/Organization \\
\hline
Number of URL/@/\# \\
\hline
\end{tabular}
\end{table}

\subsection{Baselines}
We compare our proposed method on the image credibility analysis task with a variety of other baselines, including two methods based on low-level features: existing information credibility analysis based only on textual content and traditional bag of visual word features (BoVW), and three types of transfer learning methods: data transfer, feature transfer and model transfer.

\begin{table*}
\centering
\caption{Performance Comparisons on the Target Testing Set}
\label{table_3}
\begin{tabular}{|l|l|l|l|l|l|l|l|l|l|}
\hline
& & \multirow{2}{*}{Method}& \multirow{2}{*}{ Accuracy} & \multicolumn{3}{|c|}{Fake Images} & \multicolumn{3}{|c|}{Real Images} \\
\cline{5-10}
& & & & Precision & Recall & $F_1$ &  Precision & Recall & $F_1$ \\
\hline
\multirow{2}{*}{Low-level Feature} & (1) &Text-based \cite{gupta2013faking} & 0.672 & 0.682 & 0.649 & 0.665 & 0.662 & 0.695 & 0.678 \\
\cline{2-10}
& (2) & BoVW \cite{zhang2009descriptive} & 0.685  & 0.685  & 0.691 & 0.689 & 0.685 & 0.679 & 0.682 \\
\hline
Data Transfer & (3) & proposed auxiliary set & 0.66 & 0.664 & 0.653 & 0.658 & 0.656 & 0.667 & 0.661 \\
\hline
\multirow{2}{*}{Feature Transfer}& (4) &ImageNet set \cite{krizhevsky2012imagenet}& 0.719 & 0.73 & 0.699 & 0.714 & 0.709 & 0.739 & 0.724 \\
\cline{2-10}
& (5) & proposed auxiliary set & 0.741 & 0.747 & 0.733 & 0.74 & 0.735 & 0.749 & 0.742 \\
\hline
\multirow{2}{*}{Model Transfer}& (6) &ImageNet set & 0.745 & 0.758 & 0.723 & 0.74 & 0.732 & 0.767 & 0.75 \\
\cline{2-10}
& (7) & proposed auxiliary set & 0.755 & 0.771 & 0.728 & 0.749 & 0.74 & 0.782 & 0.761 \\
\hline
Iterative Transfer & (8) & iterative transfer learning & \textbf{0.771} & \textbf{0.785} & \textbf{0.75}  & \textbf{0.767}  & \textbf{0.758} & \textbf{0.792} & \textbf{0.775}  \\
\hline
\end{tabular}
\end{table*}

\begin{itemize}
  \item
  \emph{Text-based features } Existing studies on image credibility analysis are mostly based on features extracted from the surrounding text of images \cite{gupta2013faking} \cite{jin2015}. Following these studies, we extract 16 text-based features (Table \ref{table_2}) from the tweet text content of images.
  \item
  \emph{Bag-of-Visual-Words features (BoVW)} The BoVW model has been successfully employed in many computer vision applications \cite{zhang2009descriptive}. In BOVW, the local features are extracted from each image. A vocabulary of visual words is constructed via clustering algorithms. Each image is represented as a visual document consisting of a set of visual words. Here, BoVW features are extracted based on SIFT descriptors \cite{lowe1999object}.

  \item
  \emph{Data Transfer} We train a CNN with only auxiliary image set and then directly apply it to predict the labels of images in the target testing set. This result would indicates how closely the auxiliary set relates to the target task.
  \item
  \emph{Feature Transfer} Features learned with a deep CNN on large scale datasets can smoothly generalize to related problems \cite{donahue2013decaf} \cite{razavian2014cnn}. To transfer features from the source domain to target domain, we could use feature vectors extracted from different layers of the network: the last convolution layer ($C5$) and the fully connected layers ($FC6$ and $FC7$). We extract the $C5$ feature after the max-pooling layer to reduce the feature dimension. The output feature of $C5$ has 9216 dimensions and the $FC6,7$ feature has 4096 dimensions. For performance comparisons, we adopt features transferred from two different source networks. Both networks have the same structure and hyper-parameter settings but one is trained on the general ImageNet dataset, the other is trained on our proposed auxiliary image set.
  \item
  \emph{Model Transfer}
  In data transfer and feature transfer learning, we train a CNN with only the auxiliary set. On one hand,  labels in the auxiliary set is weakly-labeled and noisy, labels in the target training set all have confident ground-truth. We should utilize the information in the training set. On the other hand, the training set has only 13,154 images, the fully-supervised deep network will generally overfit the training data. In order to train a robust classifier, we transfer weights learned with only the auxiliary set to the target task. We remove the output layer $FC8$ of the pre-trained network and add an adaptation layer $FC8'$ as the replacement. The parameters of the five convolutional layers $C1-C5$, and two fully connected layers $FC6$ and $FC7$ are first trained on the auxiliary set, then transferred to the target task. Finally, we train the new network with confidently labeled data in the target training set. For comparison, ImageNet set and our proposed auxiliary image set are used for initial parameter learning respectively.
\end{itemize}

\subsection{Performance Evaluation}
We give a detailed performance comparison of different methods for the image credibility analysis task in Table \ref{table_3}. Descriptions of the methods involved in this table are as follows:

\begin{itemize}
  \item
  Row (1)-(2) are Logistic Regression classification methods based on low-level features: 16 text-based features in Table \ref{table_2} (row (1)) and BoVW features (Row (2)).
  \item
  Row (4) and (5) are feature transfer methods from general ImageNet set and our proposed auxiliary set respectively. For both methods, we use the output of the $FC6$ layer in corresponding network as transferred features. The results are also gained through Logistic Regression.
  \item
  Row (6) and (7) are model transfer methods from general ImageNet set and our propsed auxiliary set respectively.
\end{itemize}

Generally, the domain transferred methods proposed in this paper achieve very promising results for image credibility analysis. Features transferred from our pre-trained CNN model with our proposed auxiliary image domain (Row (5)) outperforms all the three baseline methods, including text-based features, BoVW features and ImageNet Features. Our model transfer method based on proposed auxiliary set (Row (7)) also beats general ImageNet and other baselines. Moreover, our proposed iterative transfer learning method (Row (8)) achieves the best performance among others and significantly boost the accuracy to 77.1\%, which is about 10\% higher than traditional text-based method. This result validates the effectiveness of our proposed approach for the image credibility analysis task.

Upon a closer look at each result, we can also draw several interesting and important conclusions from three comparisons:

\textbf{Visual Content v.s. Textual Content}
Visual content is very important for image credibility analysis. Comparing the results of the text-based method with the other visual-based methods, the accuracy performance has been boosted from 0.672 to 0.77. The hand-crafted BoVW features performs slightly better than text-based features, as they are not able to effectively capture complicated semantics of fake images. However, with CNN, the visual appearance of images prove to be very essential for analyzing image credibility.

\begin{figure}
\centering
\includegraphics[width=3in]{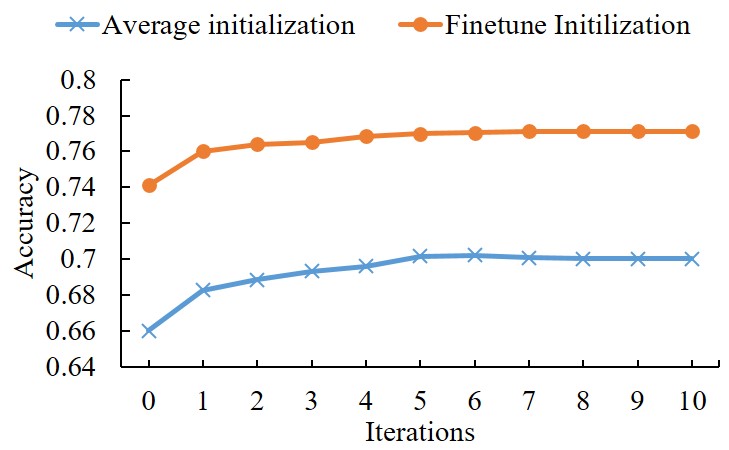}
\caption{Comparison of two weight initialization methods.}
\label{fig_6}
\end{figure}

\textbf{CNN v.s. Traditional Features}
Representation learning with CNN outperforms traditional hand-crafted features. With low-level SIFT features representing images, the BoVW method outputs similar results to the text-based method. However, the accuracy gains more than 3\% when we use the features extracted from the $FC6$ layer of two pre-trained CNN (Row (4),(5)). This result indicates the good representation learning ability of CNN and validate our choice of this model to achieve the state-of-the-art performance.

With parameters learned in the original CNN on the auxiliary set and a few confidently labeled target data for fine-tuning, the transfer learned CNN (Row (7)) achieves even better results than the direct feature transfer methods. This means that CNN can learn more semantics than general shallow learning algorithms.

\textbf{Related Domain v.s. General Domain}
A related source auxiliary domain is crucial for transfer learning. Without any information from the target domain, our data transfer method achieves the accuracy of 0.66, which is much higher than random prediction. This indicates our collected auxiliary set is quite related to the image credibility analysis task. For feature transfer and model transfer, the advantage of our proposed auxiliary set is apparent, especially for feature transfer: the result with auxiliary set (Row (5)) is 2\% higher than that of ImageNet (Row (4)). Moreover, the iterative transfer learning with proposed auxiliary set is 0.77, which outperforms the best ImageNet-based result (Row (6)) by 2.5\%. Considering that the ImageNet set is a very large image set widely used for transfer learning in many visual recognition tasks and it covers 1000 categories and has accurate human-annotated labels, our machine-generated auxiliary set is quite successful for the image credibility task. This result validates the effectiveness of our proposed method for automatically construction of the related auxiliary image set.

\subsection{Performance of Iterative Transfer Learning}

\subsubsection{Weight Initialization}

The first step of iterative transfer learning is assigning initial weights to both the auxiliary set and target training set. Our collected auxiliary set is about 50 times larger than the target training set, with traditional average weighting initialization, the impact of the smaller target dataset could be easily overwhelmed by the auxiliary set. We propose to use the result of model transfer learning to initialize the weight of data in the auxiliary set. To be specific, we first train a fine-tuned model (Row (6) in Table \ref{table_3}), then the predictions of this model for the auxiliary set is used as its initial weights: accurately predicted instances have larger weights than incorrectly predicted ones.

In Figure \ref{fig_6}, we plot the accuracy at each iteration for these two initialization methods. We can observe that the performance of average initialization is much worse than that of fine-tune initialization. Moreover, our iterative transfer learning can converge in just a few iterations: the result of fine-tune initialization stays at the best accuracy stably in just five iterations.

\subsubsection{Patterns in fake images}

\begin{figure}
\centering
\includegraphics[width=3in]{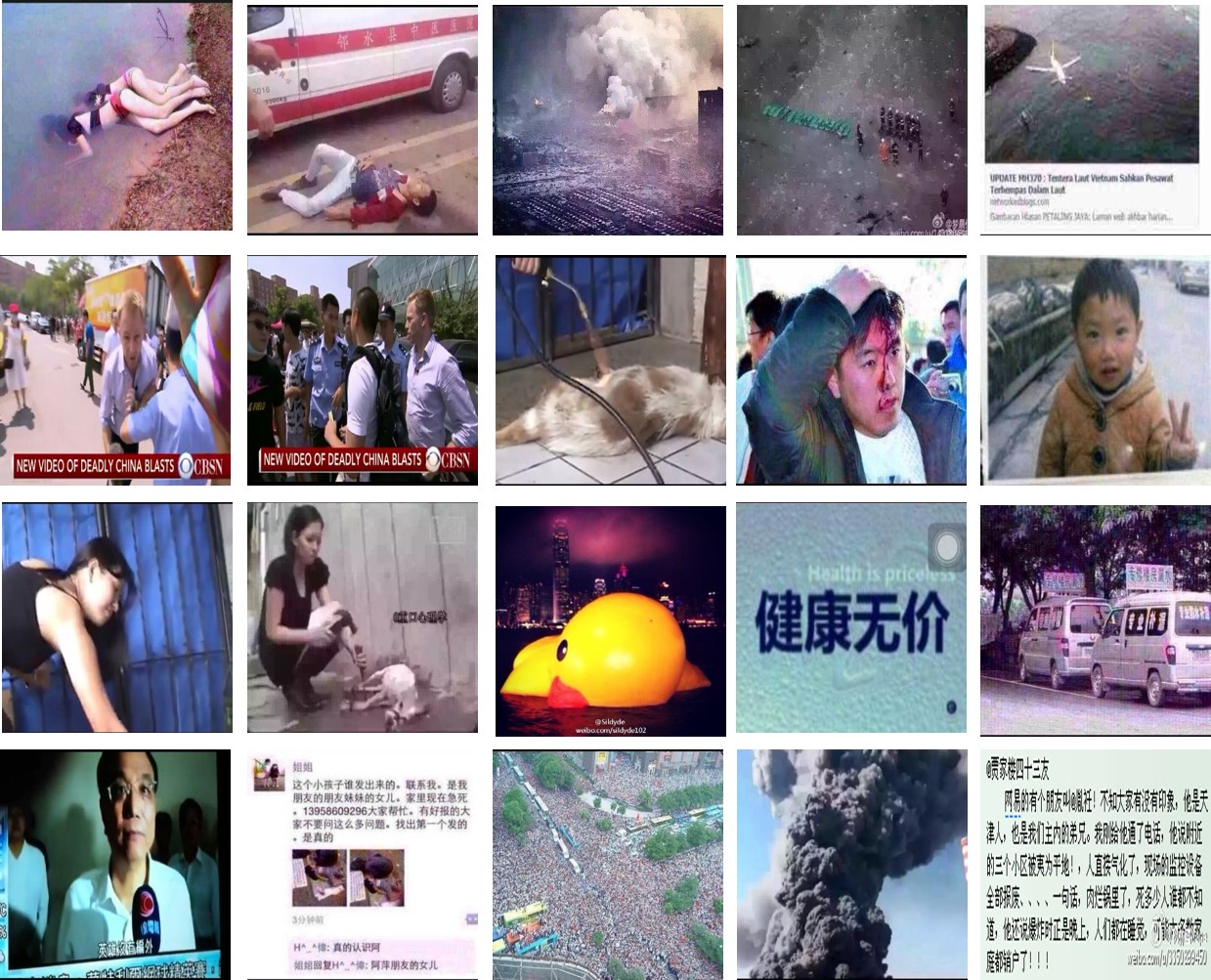}
\caption{Top ranked fake images by iterative transfer learning.}
\label{fig_7}
\end{figure}

In addition to the overall accuracy, we are very interested in understanding how our model works. To illustrate the visual patterns learned by our proposed iterative transfer learning model, we carefully examine images predicted correctly by this model. Figure \ref{fig_7} shows 15 top ranked fake images accurately predicted by this model. We can observe that our proposed method can detect fake images in a wide variety of events. Most fake images depict disturbing events, such as an accident, abuse, injury and conflict. Moreover, these top ranked fake images have some very interesting patterns: they are visually striking and likely to invoke negative emotions. They attract viewers attention on the themed events and stimulate their negative emotions. This feature would lead to wider spread of the fake images.

From this illustration, our model seem to capture accurate and complicated fake image patterns. Apparently, these patterns are quite different from the patterns used to recognize common types of objects or normal events. They are more subtle and complicated than their direct visual appearance. With such less object-specific patterns, our model are expected to generalize to fake images in various unseen events.

\subsection{Performance of Feature Transfer}
A deep network can easily overfit on a small dataset. In the cases where high-quality labeled data are rare, the feature transfer method is still a good choice as shown below.

\begin{table}
\centering
\caption{Accuracy Results of Feature Transfer}
\label{table_4}
\begin{tabular}{|c|c|c|c|c|}
\hline
\multirow{2}{*}{Dataset for Training CNN}& \multicolumn{3}{|c|}{Features}\\
\cline{2-4}
 & $C5$ & $FC6$ & $FC7$ \\
\hline
ImageNet set & 0.701 & 0.719 & 0.703 \\
\hline
Auxiliary set & 0.748 & 0.748 & 0.755  \\
\hline
\end{tabular}
\end{table}

Features extracted from different levels of convolution neural networks have different abstractions of image semantics. In this section, we perform experiments with features extracted from the fifth convolution level ($C5$) and two fully-connected layers ($FC6$ and $FC7$) from CNN trained with the ImageNet set and our auxiliary set, respectively.

As illustrated in Table \ref{table_4}, unsurprisingly, the results of our proposed source CNN outperform AlexNet for all layers. For each model, the performance variation is small. The best result based on ImageNet set comes from features extracted from $FC6$ layer, and for auxiliary set comes from $FC7$ layer. The structure of CNN determines that the higher layer represents higher abstraction and more target-specific features. As our auxiliary set is directly related to target task, the $FC7$ layer of CNN trained on it captures direct semantics for image credibility analysis and results in the best performance in this model.

\subsection{Failure Cases}
\begin{figure}[!t]
\centering
\subfigure[False positive images]{

\includegraphics[width=3in]{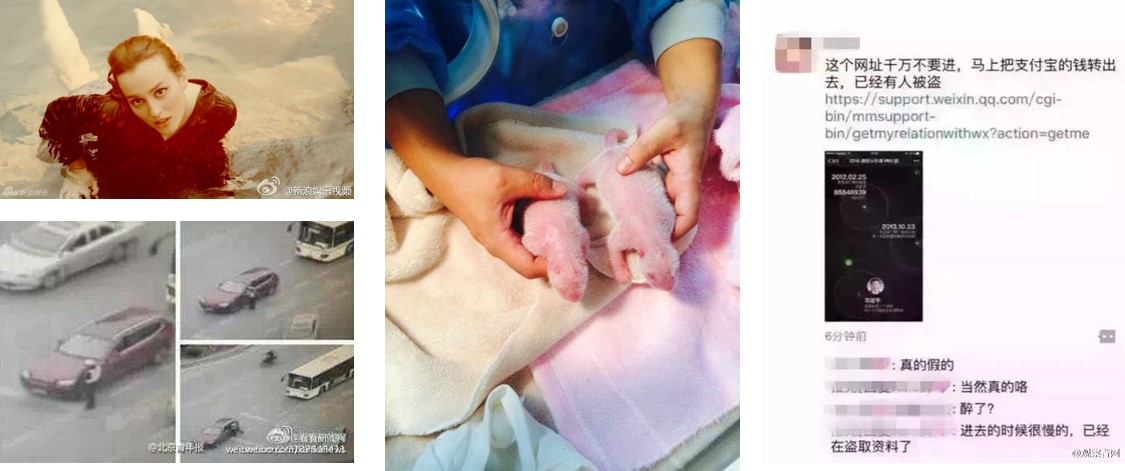}
\label{fig_8a}
}
\subfigure[False negative images]{
\includegraphics[width=3in]{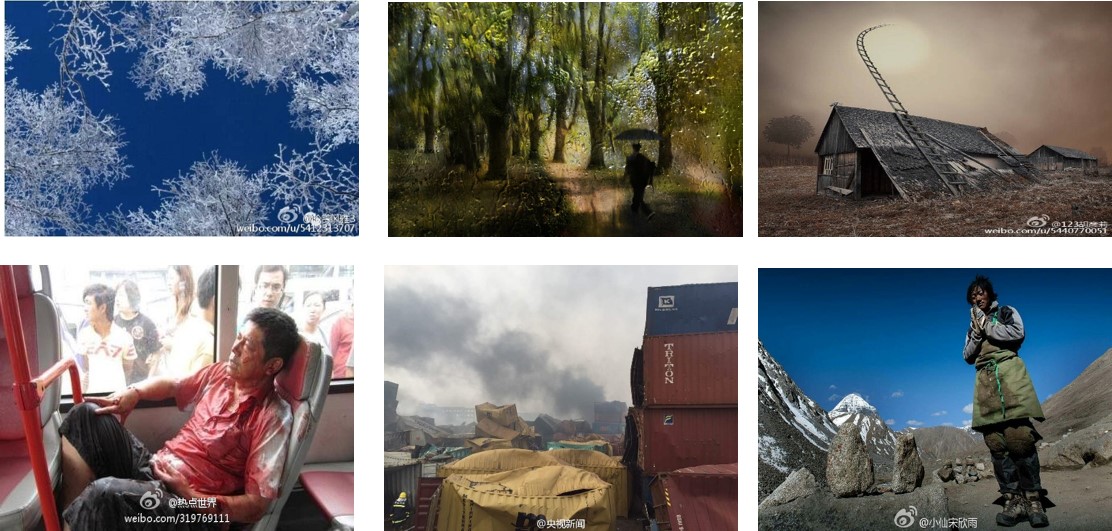}
\label{fig_8b}
}

\caption{Top ranked incorrectly predicted images by iterative transfer learning.}
\label{fig_8}
\end{figure}

After analyzing the correctly detected images, we investigate the incorrect cases of our model by examining incorrectly predicted images. This analysis reveals the weakness of our model and can help us  further improve the method. There are two kinds of failure cases. False positive cases (false alarms) are real images mistakenly classified as fake ones. Conversely, false negative cases are fake images mistakenly classified as real images. We show the top ranked incorrectly predicted images of both types in Figure \ref{fig_8}.

It is revealed in these cases that our model actually captures general fake image patterns. Incorrectly predicted real images have the same visual patterns of general fake images. They tend to be eye-catching and visually striking, just as fake images in Figure \ref{fig_7}. For these incorrectly predicted cases, the patterns learned in our model would perform poorly in detecting them accurately. In the future, we may consider incorporate information beyond the image content to further improve the overall prediction using multimodal information.

\section{conclusions and Future Work}
Image credibility analysis is a challenging and interesting problem. It requires the comprehension of complicated visual semantics in fake images from various events. In this paper, we propose a domain transferred deep convolutional neural networks to solve this problem. First, we collect a large scale weakly-labeled fake image dataset from social media as the auxiliary set for transfer learning. We then build a CNN to learn with weighted instances. Finally, we employ an AdaBoost-like algorithm to assign weights to instances in both the auxiliary and target datasets during the CNN learning. This iterative transfer learning produces a reliable CNN model to detect fake images.

The experimental results on a 40k real-word image set suggest that our proposed domain transferred CNN generates promising results for the image credibility analysis task. It outperforms several competing baselines, including a traditional text-based method, low-level BoVW features, and three typical transfer learning methods (data transfer, feature transfer and model transfer). Moreover, our proposed model reveals interesting general patterns of fake images: they tend to be low quality, event-centric, more eye-catching and disturbing than real ones. With these complicated semantics, our model can detect fake images in various events. We attribute the success of our approach to the incorporation of two main aspects: the successful construction of a large scale auxiliary set directly related to our task and the iterative transfer learning to incorporate the representation learning ability of CNN and valuable information in the auxiliary and target sets.

In the future, we plan to fuse textual content, visual content and social context for analyzing  information credibility on social media. We already find that textual content is a good indicator for image credibility and has a great contribution on constructing the auxiliary domain dataset. We believe that a multi-modality model would integrate all the knowledge from heterogeneous aspects and further improve the performance of online information credibility analysis.

\ifCLASSOPTIONcaptionsoff
  \newpage
\fi


\bibliographystyle{IEEEtran}

\bibliography{ref}

\vfill

\end{document}